\documentclass[aps,prb,groupedaddress,reprint,floatfix,showpacs, nobalancelastpage
]{revtex4-1}
\usepackage{graphicx,amsmath,amssymb,bm,hyperref,color,engord}
\usepackage{epstopdf}
\definecolor{blue}{rgb}{0.0,0.0,1.0}
\definecolor{black}{rgb}{0.0,0.0,0.0}
\definecolor{red}{rgb}{1.0,0.0,0.0}
\hypersetup{colorlinks=true,breaklinks, linkcolor=blue,urlcolor=blue,citecolor=red}
\begin{document}

\title{Theory of correlated electron transport and inelastic  tunneling spectroscopy}
\author{Kelly R. Patton}
\email[\hspace{-1.4mm}]{kpatton@lsu.edu}
\affiliation{Department of Physics and Astronomy, Louisiana State University, Baton Rouge, Louisiana 70803 }
\date{\today}
\begin{abstract}
For a non-superconducting system, the electronic tunneling current through an insulating barrier is calculated, including interaction effects.   The exact Hamiltonian of the full system is projected onto the subspaces of the ``left'' and ``right'' leads.  In the weak tunneling limit  the well-known tunneling Hamiltonian is recovered, along with an additional term. This additional term originates from the projection of the electron-electron interaction onto each subsystem and corresponds to correlated tunneling.  It is shown that the tunneling current is determined by---in addition to the single-particle density of states---the  spin-spin and density-density susceptibilities.   The signatures of which  have recently been observed in several  experiments.     
\end{abstract}
\maketitle
\section{\label{sec: intro}Introduction}
Tunneling experiments  were one of the first major confirmations of the Bardeen-Cooper-Schrieffer (BCS) theory of superconductivity.\cite{GiaeverPRL60}     These experiments  showed a dramatic suppression in the conductance near the Fermi energy.   The theoretical explanation  of this was the formation of a gap in the quasi-particle density of states of the superconductor. \cite{BardeenPRL61}  This essentially began the relationship linking a tunneling current in a many-body system to the single-particle density of states. 

To describe tunneling at a general many-body level, Cohen {\it et al.}\cite{CohenPRL62}\ later introduced the tunneling or transfer Hamiltonian, consisting of two disjoint, fully interacting Hamiltonians connected by a tunneling term. This tunneling term  removes electrons from one system and creates them in the other. Although phenomenological, the tunneling Hamiltonian approach has become and remains the de facto standard for theoretically  interpreting tunneling experiments in condensed matter, across a wide range of systems.   Several attempts have been made  to put many-body tunneling, with or without the tunneling Hamiltonian,  on more rigorous theoretical ground,\cite{PrangePR63,*ZawadowskiPR67,*AppelbaumPR69,*CaroliJPhysC71a,*FeuchtwantPRB74a} but owing to the inherent difficulties of this problem, it still remains open. 

Nevertheless, the standard tunneling formalism has been used extensively to  account for an extraordinarily  wide variety of experiments.  Although, with the ever increasing refinement of experimental techniques, experiments are starting to probe physics not captured by this formalism.    One area where modifications are needed is in the regime of inelastic tunneling, where the tunneling electron loses energy through interactions with the barrier or other additional degrees of freedom of the system being probed.  These effects are typically accounted for by simply adding additional transfer terms.\cite{PerssonPRL87}    Inelastic electron tunneling spectroscopy (IETS), most notably scanning tunneling microscopy IETS (STM-IETS),  has be used to probe the internal vibrational modes of molecular adsorbates on surfaces.  More recent experiments \cite{HeinrichScience04,HirjibehedinScience06,BalashovPRL06}  on single magnetic atoms, finite spin chains, and magnetic substrates have shown evidence of spin related inelastic tunneling.  Here, especially for the adatom chains,  the current-voltage (I-V) curves show features where one would expect spin excitations of the system, such as the singlet-triplet excitation of neighboring pairs.   Explaining such characteristics is outside the scope of the standard tunneling approach,  which relates an I-V curve, or conductance, to single-particle properties, namely the local single-particle density of states.  Spin excitations, such as the singlet-triplet transition,  are manifestly two-particle properties, related to the spin (density) susceptibility or spin-spin correlation function.  Several authors have been able to account for the experimental results of these experiments  remarkably well,\cite{RossierPRL09, FranssonNanoLett09} by introducing an interaction effect between the STM tip and adatoms, leading to so-called spin mediated tunneling.  The microscopic origin of such a term and how it should be generalized, to say other spin excitations such as magons,\cite{BalashovPRL06}  isn't immediately  clear though.   In principle these and other interaction effects should occur in virtually all tunneling experiments.\cite{AppelbaumPR67}   One would like a theoretical foundation  that is  first principles driven such that it could be applied to a wide variety of systems and also be generalizable to others in a systematic manner.  The main goal of this article is to give such a derivation of the tunneling current, in the spirit of the transfer Hamiltonian,  that includes these interaction or two-body effects. 

In the next section we start with a general interacting Hamiltonian for a system with a tunneling barrier.  The original Hamiltonian is then projected onto the low-energy (below the height of the barrier) ``left'' and ``right''  subspaces.  This projection is conjectured to capture the most salient tunneling processes.    From which, the standard tunneling Hamiltonian is recovered along with an additional term, which describes correlated or interaction meditated tunneling, stemming from the original electron-electron interaction.  In Sec.\ \ref{sec:current} we calculate the steady-state tunneling current through the barrier  with respect to this effective tunneling Hamiltonian.  The result gives an additional contribution to the current from two-particle susceptibilities,  as well as the well-known contribution determined by the single-particle density of states.

\section{\label{sec: Hamiltonian}Derivation of effective tunneling Hamiltonian}
\begin{figure}
\includegraphics[width=.48\textwidth]{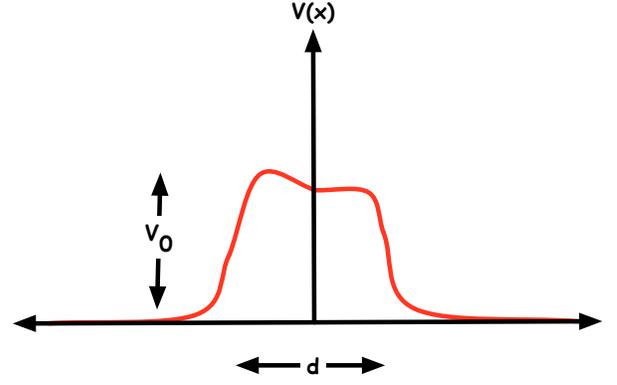}%
\caption{Schematic of a tunneling barrier for the full Hamiltonian, Eq.~\eqref{exact Hamiltonian}, having a characteristic width $d$ and height $V_{0}$.  \label{fig1}}
\end{figure}

Here we consider a one-dimensional system for notational convenience; the generalization to higher dimensions is straightforward. 
For an interacting system in the presence of a tunneling barrier, the  grand-canonical Hamiltonian of the {\it entire} system is given in second quantization by (setting $\hbar=1$)
\begin{align}
\label{exact Hamiltonian}
&H=\sum_{\sigma}\int dx\, \Psi^{\dagger}_{\sigma}(x)\Big[-\frac{\nabla^{2}}{2m}-\mu+V(x)\Big]\Psi^{}_{\sigma}(x)\nonumber\\+&\frac{1}{2}\sum_{\sigma,\sigma'}\int dx dx'\,\Psi^{\dagger}_{\sigma}(x)\Psi^{\dagger}_{\sigma'}(x')U(x,x')\Psi^{}_{\sigma'}(x')\Psi^{}_{\sigma}(x),
\end{align}
where $U(x,x')$ is the electron-electron interaction (assumed to be symmetric), $\mu$ is the chemical potential, and  $V(x)$ is taken to be the tunneling barrier, but could also contain disorder, lattice fields or other single-particle potentials, see Fig.\ \ref{fig1}.   The field operators $\Psi(x)$ and $\Psi^{\dagger}(x)$ obey the standard fermionic anticommuation relations $\{\Psi_{\sigma}(x),\Psi_{\sigma'}(x')\}=0$, $\{\Psi^{}_{\sigma}(x) ,\Psi^{\dagger}_{\sigma'}(x')\}=\delta_{\sigma,\sigma'}\delta(x-x')$, and
can be expressed  in terms of mode creation and annihilation operators by
\begin{align}
\Psi^{}_{\sigma}(x)&=\sum_{k}\psi^{}_{k\sigma}(x)a^{}_{k\sigma}\\
\Psi^{\dagger}_{\sigma}(x)&=\sum_{k}\psi^{*}_{k\sigma}(x)a^{\dagger}_{k\sigma},
\end{align}
where $\{a^{}_{k\sigma},a^{}_{k'\sigma'}\}=0$, $\{a^{}_{k\sigma},a^{\dagger}_{k'\sigma'}\}=\delta_{\sigma,\sigma'}\delta_{k,k'}$,
and $\psi^{}_{k\sigma}(x)$ are the exact single-particle eigenstates of \eqref{exact Hamiltonian}.  

\begin{figure}
\includegraphics[width=.48\textwidth]{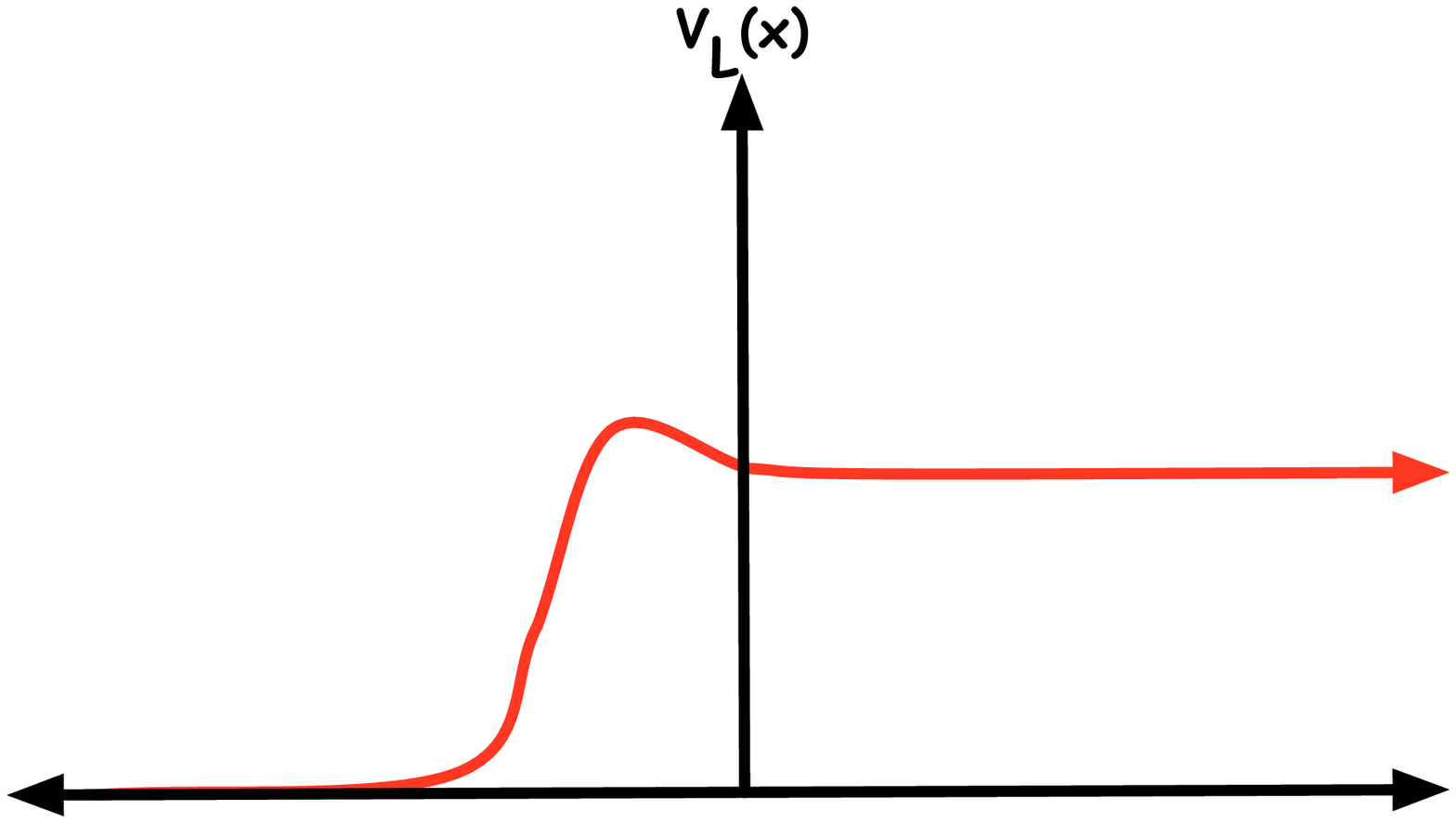}%
\caption{\label{fig2}Potential barrier for $H_{\rm L}$, defined in \eqref{VL}.}
\end{figure}

\begin{figure}
\includegraphics[width=.48\textwidth]{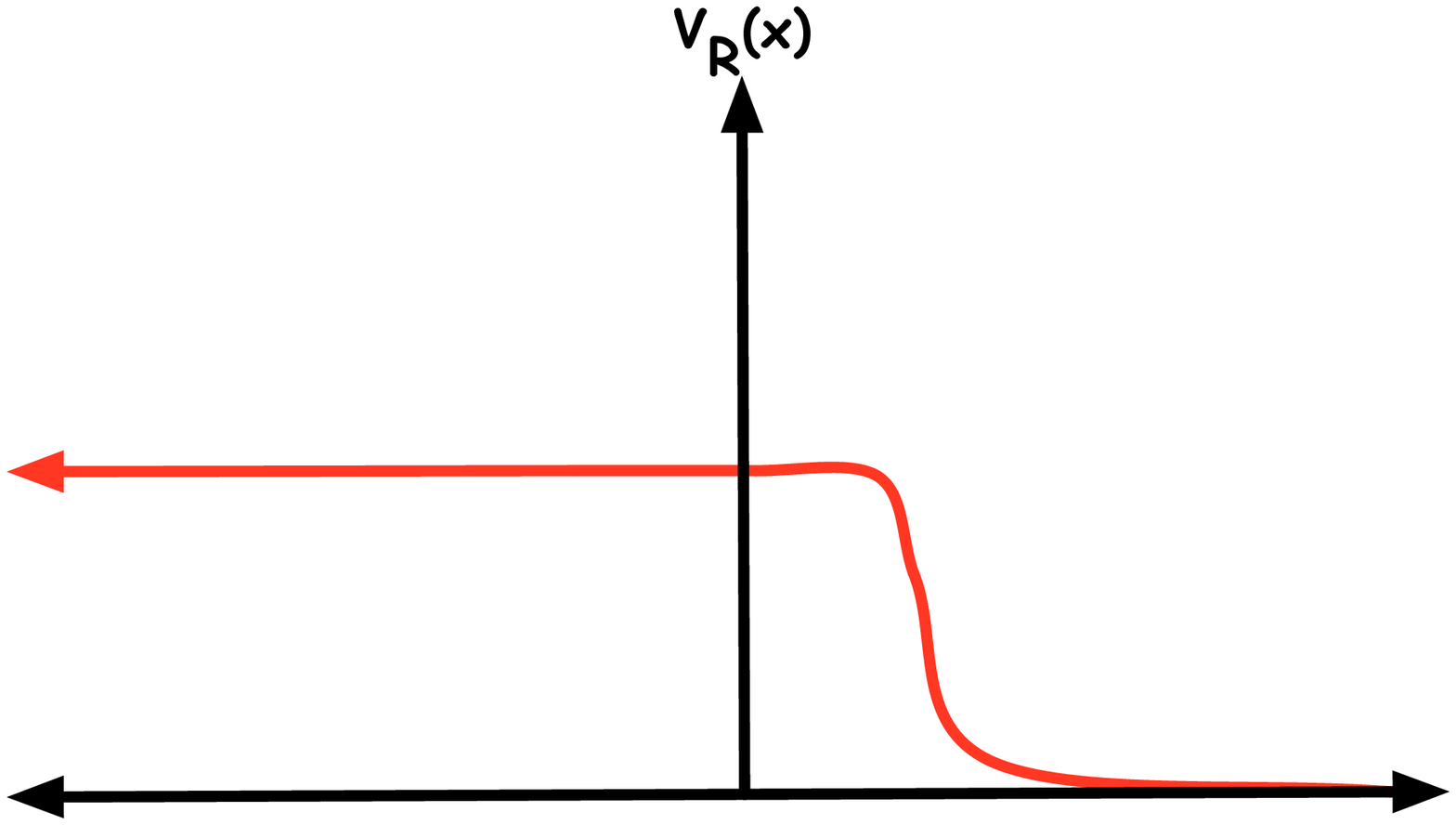}%
\caption{\label{fig3} Potential barrier for $H_{\rm R}$, defined in \eqref{VR}. }
\end{figure}

Next, we consider two related Hamiltonians 

\begin{align}
\label{left Hamiltonian}
&H_{\rm L}=\sum_{\sigma}\int dx\, \Psi^{\dagger}_{{\rm L},\sigma}(x)\Big[-\frac{\nabla^{2}}{2m}-\mu+V_{\rm L}(x)\Big]\Psi^{}_{{\rm L},\sigma}(x)\nonumber\\+&\frac{1}{2}\hspace{-0.5mm}\sum_{\sigma,\sigma'}\hspace{-0.5mm}\int\hspace{-1mm} dx dx'\Psi^{\dagger}_{{\rm L},\sigma}(x)\Psi^{\dagger}_{{\rm L},\sigma'}(x')U(x,x')\Psi^{}_{{\rm L},\sigma'}(x')\Psi^{}_{{\rm L},\sigma}(x),
\end{align}
and
\begin{align}
\label{right Hamiltonian}
&H_{\rm R}=\sum_{\sigma}\int dx\, \Psi^{\dagger}_{{\rm R},\sigma}(x)\Big[-\frac{\nabla^{2}}{2m}-\mu+V_{\rm R}(x)\Big]\Psi^{}_{{\rm R},\sigma}(x)\nonumber\\ +&\frac{1}{2}\hspace{-0.5mm}\sum_{\sigma,\sigma'}\hspace{-0.5mm}\int\hspace{-1mm} dx dx'\Psi^{\dagger}_{{\rm R},\sigma}(x)\Psi^{\dagger}_{{\rm R},\sigma'}(x')U(x,x')\Psi^{}_{{\rm R},\sigma'}(x')\Psi^{}_{{\rm R},\sigma}(x),
\end{align}

where
\begin{equation}
\label{VL}
V^{}_{\rm L}(x)=\begin{cases}
V(x) ,&\text{for~} x\leq a\,\\ V(a), & \text{for~} x\geq a\,,
\end{cases}
\end{equation}
and
\begin{equation}
\label{VR}
V^{}_{\rm R}(x)=\begin{cases}
V(x), &\text{for~} x\geq a\,\\ V(a), & \text{for~} x\leq a.
\end{cases}
\end{equation}
The parameter $a$ is an arbitrary point within the barrier, such that $V(x)=V_{\rm L}(x)\Theta(a-x)+V_{\rm R}(x)\Theta(x-a)$, where $\Theta(x)$ is the Heavyside step function; see Figs.\ \ref{fig2} and \ref{fig3}.
The field  operators in \eqref{left Hamiltonian} and \eqref{right Hamiltonian}  are given by 
\begin{align}
\Psi^{}_{\rm L,\sigma}(x)&=\sum_{k}\psi^{}_{{\rm L},k\sigma}(x)a^{}_{{\rm L},k\sigma}\\
\Psi^{}_{\rm R,\sigma}(x)&=\sum_{k}\psi^{}_{{\rm R},k\sigma}(x)a^{}_{{\rm R},k\sigma},
\end{align}
where $\psi^{}_{{\rm L},k\sigma}$ and $\psi^{}_{{\rm R},k\sigma}$ are the exact and complete single-particle eigenstates of $H_{\rm L}$ and $H_{\rm R}$ respectively.   

The Hilbert space of $H$, $\cal H$ can be decomposed  into the Hilbert spaces of $H_{\rm L}$ and $H_{\rm R}$, 
\begin{equation}
{\cal H}={\cal H}_{\rm L}+{\cal H}_{\rm R}+\delta{\cal H},
\end{equation}
where $\delta{\cal H}$ allows for states outside of either ${\cal H}_{\rm L}$ or ${\cal H}_{\rm R}$, e.g. resonant states within the barrier itself.  Such possibilities will not be considered here. 
One can then define projection operators
\begin{align}
P_{\rm L}&=\sum_{\alpha}|\psi_{{\rm L},\alpha}\rangle\langle\psi_{{\rm L},\alpha}|\nonumber\\P_{\rm R}&=\sum_{\alpha}|\psi_{{\rm R},\alpha}\rangle\langle\psi_{{\rm R},\alpha}|,
\end{align}
 which project the original Hilbert space on to the left and right subspaces respectively. We assume that the sum of projection operators, to a good approximation, forms a resolution of the identity in ${\cal H}$, i.e. 
\begin{equation}
P^{}_{\rm L}+P^{}_{\rm R}\simeq\openone. 
\end{equation}
The electron field operator of \eqref{exact Hamiltonian} can therefore  be written as
\begin{align}
\Psi_{\sigma}(x)&=(P^{\dagger}_{\rm L}+P^{\dagger}_{\rm R})\Psi_{\sigma}(x)(P^{}_{\rm L}+P^{}_{\rm R})\nonumber\\&=\tilde{\Psi}_{\rm L}(x)+\tilde{\Psi}_{\rm R}(x)+\delta\Psi(x),
\end{align}
where $\delta\Psi(x)$ is of order of the tunneling amplitude, i.e.,  for states below the barrier height $\delta\Psi(x)\sim O(T_{\alpha,\alpha'}=\langle\psi_{{\rm L},\alpha}|V|\psi_{{\rm R},\alpha'}\rangle)$.  $\delta\Psi(x)$ will be neglected from here on, as it will ultimately correspond to higher order terms, in the tunneling amplitude, in a calculation of the current.  
Projecting the field operator of the original  Hamiltonian onto the left and right subspaces
\begin{align}
\tilde{\Psi}_{\rm L,\sigma}(x)&=P^{\dagger}_{\rm L}\Psi_{\sigma}(x)P^{}_{\rm L}=\sum_{k}\psi^{}_{k\sigma}(x)a^{}_{\rm L,k\sigma}\nonumber\\
\tilde{\Psi}^{\dagger}_{\rm L,\sigma}(x)&=P^{\dagger}_{\rm L}\Psi^{\dagger}_{\sigma}(x)P^{}_{\rm L}=\sum_{k}\psi^{*}_{k\sigma}(x)a^{\dagger}_{\rm L,k\sigma},
\end{align}
and
\begin{align}
\tilde{\Psi}_{\rm R,\sigma}(x)&=P^{\dagger}_{\rm R}\Psi^{}_{\sigma}(x)P^{}_{\rm R}=\sum_{k}\psi^{}_{k\sigma}(x)a^{}_{\rm R,k\sigma}\nonumber\\
\tilde{\Psi}^{\dagger}_{\rm R,\sigma}(x)&=P^{\dagger}_{\rm R}\Psi^{\dagger}_{\sigma}(x)P^{}_{\rm R}=\sum_{k}\psi^{*}_{k\sigma}(x)a^{\dagger}_{\rm R,k\sigma},
\end{align}
where
$a^{}_{\rm L,k\sigma}=\int dx \,\psi^{}_{{\rm L},k\sigma}(x) \Psi^{}_{\sigma}(x)$, $
a^{\dagger}_{\rm L,k\sigma}=\int dx \,\psi^{*}_{{\rm L},k\sigma}(x) \Psi^{\dagger}_{\sigma}(x)$, $a^{}_{\rm R,k\sigma}=\int dx \,\psi^{}_{{\rm R},k\sigma}(x) \Psi^{}_{\sigma}(x)$, and $a^{\dagger}_{\rm R,k\sigma}=\int dx \,\psi^{*}_{{\rm R},k\sigma}(x) \Psi^{\dagger}_{\sigma}(x).$
From which it readily follows that $\big\{a^{}_{\rm L,k\sigma},a^{\dagger}_{\rm L,k'\sigma'}\big\}=\big\{a^{}_{\rm R,k\sigma},a^{\dagger}_{\rm R,k'\sigma'}\big\}=\delta_{\sigma,\sigma'}\delta_{k,k'}$, $\big\{a^{}_{\rm L,k\sigma},a^{}_{\rm L,k'\sigma'}\big\}=\big\{a^{}_{\rm R,k\sigma},a^{}_{\rm R,k'\sigma'}\big\}= \big\{a^{}_{\rm L,k\sigma},a^{}_{\rm R,k'\sigma'}\big\}=0$, and  $\big\{a^{}_{\rm L,k\sigma},a^{\dagger}_{\rm R,k'\sigma'}\big\}=\big<\psi^{}_{{\rm L},k\sigma}|\psi^{}_{{\rm R},k'\sigma'}\big>$. 
In general the left and right states are not orthogonal; $\big<\psi^{}_{{\rm L},\alpha}|\psi^{}_{{\rm R},\alpha'}\big>\neq 0$.  For states with energy below the barrier height $\big<\psi^{}_{{\rm L},\alpha}|\psi^{}_{{\rm R},\alpha'}\big>\sim O(T_{\alpha,\alpha'})$.

As the exact single-particle eigenstates of \eqref{exact Hamiltonian}, \eqref{left Hamiltonian}, and \eqref{right Hamiltonian} are assumed to form a complete basies, we can expand the original eigenfunctions  in terms of the left and right basis
\begin{align}
\psi_{k\sigma}(x)&=\sum_{k'}c^{\rm L}_{k,k'}\psi_{{\rm L},k'\sigma}(x)\\
\psi_{k\sigma}(x)&=\sum_{k'}c^{\rm R}_{k,k'}\psi_{{\rm R},k'\sigma}(x),
\end{align}
where $c^{\rm L}_{k,k'}=\big<\psi_{{\rm L},k}|\psi_{k'}\big>$ and $c^{\rm R}_{k,k'}=\big<\psi_{{\rm R},k}|\psi_{k'}\big>$. The expansion coefficients can be obtained from assuming an explicit form of the tunneling barrier \cite{PrangePR63} or within WKB (Wentzel-Kramers-Brillouin) theory  for an arbitrary potential. \cite{Benderbook}  For a wide tall barrier and for states with energy much less than the barrier height $c^{\rm L}_{k,k'} \propto\delta_{k,k'}+O(T_{k,k'})$ and $c^{\rm R}_{k,k'}\propto\delta_{k,k'}+O(T_{k,k'})$.  Thus, apart from an overall proportionality constant,  to leading order in the tunneling amplitude 
\begin{align}
\tilde{\Psi}_{{\rm L},\sigma}(x)&\simeq\Psi_{{\rm L},\sigma}(x)+O(T_{k,k'}) \\
\tilde{\Psi}_{{\rm R},\sigma}(x)&\simeq\Psi_{{\rm R},\sigma}(x)+O(T_{k,k'}).
\end{align}
Therefore, the original  fields operators, to leading order in the tunneling, can be decomposed into the left and right subsystems by  
\begin{align}
\label{approximate projected operators}
\Psi_{\sigma}(x)&\simeq\Psi_{\rm L,\sigma}(x)+\Psi_{\rm R,\sigma}(x)+O(T_{k,k'})\nonumber\\
\Psi^{\dagger}_{\sigma}(x)&\simeq\Psi^{\dagger}_{\rm L,\sigma}(x)+\Psi^{\dagger}_{\rm R,\sigma}(x)+O(T_{k,k'}).
\end{align}
It should be noted that, at this point,  because we have restricted ourselves to states below the barrier, the field operators $\Psi^{\dagger}_{\rm L}(x)$ and $\Psi^{\dagger}_{\rm R}(x)$ in Eq.\ \eqref{approximate projected operators}, no longer create (annihilate) electrons in localized delta-function states, but instead in some broaden  approximation.   We will not deal with this technicality and assume that for a high barrier, compared to the Fermi energy, to a good approximation the states below the barrier form a sufficiently complete set.

Expressing the exact Hamiltonian \eqref{exact Hamiltonian} in terms of the approximate projected operators \eqref{approximate projected operators}, and for now neglecting interactions,  gives
\begin{align}
\label{H_{0} tunneling}
H\approx&\sum_{k,\sigma}(\epsilon^{}_{k\sigma}-\mu)a^{\dagger}_{\rm L,k\sigma}a^{}_{\rm L,k\sigma}+\sum_{k,\sigma}(\epsilon^{}_{k\sigma}-\mu)a^{\dagger}_{\rm R,k\sigma}a^{}_{\rm R,k\sigma}\nonumber\\&+\sum_{k,k',\sigma}\big[T_{k,k'}a^{\dagger}_{\rm L,k\sigma}a^{}_{\rm R,k'\sigma}+{\rm H.c.}\big]\nonumber\\&+\sum_{k,k',\sigma}\big[{\xi}_{k,k',\sigma}(\epsilon^{}_{k'\sigma}-\mu)a^{\dagger}_{\rm L,k\sigma}a^{}_{\rm R,k'\sigma}+{\rm H.c.}\big],
\end{align}
where terms such as $\int dx\, \Psi^{\dagger}_{\rm L}(x)V(x)\Psi^{}_{\rm L}(x)$ that only lead to a change in the density and energies near the barrier have been neglected, and ${\xi}_{k,k'}=\big<\psi^{}_{{\rm L},k}|\psi^{}_{{\rm R},k'}\big>$.    At this level, the fact that the eigenstates of the left and right sides are not orthogonal simply leads to a renormalization of the tunneling matrix elements, $T_{k,k'}\rightarrow T_{k,k'}+{\xi}_{k,k'}(\epsilon^{}_{k'}-\mu)$.  Even in the approximation ${\xi}_{k,k'}\rightarrow0$, the important and dominate physics is stilled captured.  Thus, neglecting these terms or equivalently setting ${\xi}_{k,k'}=0$, one can still expect to obtain quantitatively correct results, of course this approximation can be relaxed. \cite{FranssonPRB01} 
Also, this simplification exactly reduces \eqref{H_{0} tunneling} to the standard tunneling Hamiltonian. 
 In addition, this greatly simplifies calculations of expectation values, i.e Green's functions, as now $\{\Psi^{}_{\rm L,\sigma}(x),\Psi^{\dagger}_{\rm R,\sigma}(x)\}=0$, which we will assume from here on.   With such an approximation we have essentially arrived back at the starting point for the standard tunneling Hamiltonian.  One may  wonder why all of the previous formalities were necessary.  For one, it demonstrates  what approximations and assumptions are made when one is using the standard tunneling Hamiltonian formalism, and secondly it will allow us to treat the effects of interactions on tunneling quantitatively, which we now turn to. 

Normally, interactions are introduced simply by adding the appropriate interactions terms to the left and right subsystems.   This neglects the cross terms generated by expressing the original operators in terms of the approximate left and right states.  As we will see, it is these cross terms that are responsible for the electronic inelastic tunneling properties.  
Expressing the interaction term of \eqref{exact Hamiltonian} in terms of the projected operators, Eqs.\ \eqref{approximate projected operators}, leads to (see the Appendix \ref{Details of the derivation of the tunneling Hamiltonian} for details) 

\begin{equation}
\label{tunneling Hamiltonian}
H\approx H_{\rm L}+H_{\rm R}+H_{\rm T},
\end{equation}
where $H_{\rm L}$, $H_{\rm R}$ are the fully interacting Hamiltonians of the left and right subsystems 
\begin{widetext}
\begin{align}
&H_{\rm L}=\sum_{\sigma}\int dx\, \Psi^{\dagger}_{{\rm L},\sigma}(x)\Big[-\frac{\nabla^{2}}{2m}-\mu\Big]\Psi^{}_{{\rm L},\sigma}(x)+\frac{1}{2}\sum_{\sigma,\sigma'}\int dx dx'\,\Psi^{\dagger}_{{\rm L},\sigma}(x)\Psi^{\dagger}_{{\rm L},\sigma'}(x')U(x,x')\Psi^{}_{{\rm L},\sigma'}(x')\Psi^{}_{{\rm L},\sigma}(x),\nonumber\\
&H_{\rm R}=\sum_{\sigma}\int dx\, \Psi^{\dagger}_{{\rm R},\sigma}(x)\Big[-\frac{\nabla^{2}}{2m}-\mu\Big]\Psi^{}_{{\rm R},\sigma}(x)+\frac{1}{2}\sum_{\sigma,\sigma'}\int dx dx'\,\Psi^{\dagger}_{{\rm R},\sigma}(x)\Psi^{\dagger}_{{\rm R},\sigma'}(x')U(x,x')\Psi^{}_{{\rm R},\sigma'}(x')\Psi^{}_{{\rm R},\sigma}(x),
\end{align}
and the tunneling or transfer part is given by
\begin{align}
\label{tunneling part}
H_{\rm T}&=\sum_{\sigma}\int dx\, \big[T(x)\Psi^{\dagger}_{{\rm L},\sigma}(x)\Psi^{}_{{\rm R},\sigma}(x)+{\rm H.c.}\big]+\sum_{\sigma}\int dxdx'\, U(x,x')\Big\{\Psi^{\dagger}_{{\rm L},\sigma}(x)\Psi^{}_{{\rm R},\sigma}(x)\big[{\hat n}_{\rm L}(x')+{\hat n}_{\rm R}(x')\big]+{\rm H.c}\Big\},
\end{align}
\end{widetext}
where $\hat{n}(x)=\sum_{\sigma}\Psi^{\dagger}_{\sigma}(x)\Psi_{\sigma}(x)$. The first term of \eqref{tunneling part} is the standard single-particle tunneling, which comes from the kinetic energy term of the original Hamiltonian.  The last term is from the interaction of the original system and is normally not considered in most treatments of tunneling.  This term involves a interaction mediated transfer of electrons, or correlated hopping in a lattice model. 
\footnote{For instance, in the extended Shubin-Hubbard  model similar correlated hopping terms are included.  See for example Ref. [\onlinecite{JPhysCVonsovsky79}] and references therein.}
Although, this term might be small, as it involves the correlated transfer  electrons, it is ultimately responsible for the inelastic tunneling properties.   
\section{\label{sec:current} Evaluation of the tunneling current}

In the presence of an applied electric field  a net current will flow through the barrier.  Here we will calculate the steady-state current in the limit of weak tunneling between the left and right sub-systems.  As usual,\cite{Mahanbook} we will assume that each subsystem is separately in thermodynamic equilibrium, where the chemical potential of each side differs only by the applied voltage $eV$. The drop in the chemical potential is assumed to occur entirely within the barrier.  Because the total particle number commutes with the effective Hamiltonian,  the current is defined to be proportional to the expectation value of the time-rate-change of the number of electrons in the left (or right) sub-system.  By Heisenberg's equation of motion $(e>0)$ the current operator is given by  
\begin{equation}
{\hat I}=-e\partial^{}_{t}{\hat N}_{\rm L}=-ie\big[{\hat N}_{\rm L},H\big]=-ie\big[{\hat N}_{\rm L},H_{\rm T}\big],
\end{equation}
where ${\hat N}_{\rm L}=\sum_{\sigma} \int dx\, \Psi^{\dagger}_{{\rm L}\sigma}(x)\Psi^{}_{{\rm L}\sigma}(x)$. Evaluating the commutator gives
\begin{widetext}
\begin{equation}
\label{current operator}
{\hat I}=-ie\sum_{\sigma}\int dx\, \big[T(x)\Psi^{\dagger}_{{\rm L},\sigma}(x)\Psi^{}_{{\rm R},\sigma}(x)-{\rm H.c.}\big]-ie\sum_{\sigma}\int dxdx'\,  U(x,x')\Big\{\Psi^{\dagger}_{{\rm L},\sigma}(x)\Psi^{}_{{\rm R},\sigma}(x)\big[{\hat n}_{\rm L}(x')+{\hat n}_{\rm R}(x')\big]-{\rm H.c}\Big\}. 
\end{equation}
\end{widetext}
The nonequilibrium expectation value of \eqref{current operator} gives the measured tunneling current.  Here we obtain this expectation within linear response theory, treating the tunneling term, $H_{\rm T}$, as the perturbation.  To leading order in the tunneling, the current is given by the thermal expectation of
\begin{equation}
\label{LR current}
I=i\int \limits_{-\infty}^{t}dt'\big<\big[H_{\rm T}(t'),{\hat I}(t)\big]\big>_{H_{0}},
\end{equation}
where ${\hat O}(t)=e^{iH_{0}t}{\hat O}e^{-iH_{0}t}$, $H_{0}=H_{\rm L}+H_{\rm R}$, and we have assumed that as $t\to -\infty $ the two sub-systems are completely decoupled, i.e.~the height of the barrier is infinitely large.

 Evaluating \eqref{LR current} for a non-superconducting system and assuming local tunneling, i.e., the wave functions of one side are spatially localized about a point $x$, as is the case for an STM, the total current can be written as (See Appendix \ref{tunneling current appendix} for a complete derivation.)
\begin{equation}
\label{current part one and two}
I=I^{}_{1}+I^{}_{2}.
\end{equation}
 The first term gives the standard expression for the tunneling current. 
\begin{align}
I^{}_{1}&=2\pi e|T|^{2}\sum_{\sigma}\int d\omega\, \rho^{}_{{\rm L},\sigma}(x,\omega+eV)\rho^{}_{{\rm R},\sigma}(x,\omega)\nonumber\\&\times\big[n^{}_{\rm F}(\omega)-n^{}_{\rm F}(\omega+eV)\big],
\end{align}
which relates the single-particle local density of states
\begin{equation}
 \rho^{}_{\sigma}(x,\omega)=-\frac{1}{\pi}{\rm Im}\, G^{\rm ret}_{\sigma}(x,\omega),
\end{equation}
  to the total current, where $G^{\rm ret}$ is the retarded single-particle Green's function and $n_{\rm F}(\omega)=(\exp(\beta\omega)+1)^{-1}$.  The second term is related to the electronic  inelastic tunneling and is in general given by 
\begin{widetext}
\begin{align}
\label{I2}
&I_{2}=\nonumber\\&e U^{2}\sum_{\sigma,\sigma'}\int\limits_{-\infty}^{\infty}d\omega d\omega'\, \rho^{}_{{\rm L},\sigma}(x,\omega+eV)\rho^{}_{{\rm R},\sigma'}(x,\omega')\Big\{\big[\chi^{\rm L}_{\overline{\sigma}',\overline{\sigma},\overline{\sigma},\overline{\sigma}'}(x,\omega'-\omega-eV)+\chi^{\rm R}_{\overline{\sigma}',\overline{\sigma},\overline{\sigma},\overline{\sigma}'}(x,\omega'-\omega)\big]n^{}_{\rm F}(\omega')\big[1-n^{}_{\rm F}(\omega+eV)\big]\nonumber\\&-\big[\chi^{\rm L}_{\overline{\sigma},\overline{\sigma}',\overline{\sigma}',\overline{\sigma}}(x,\omega-\omega'+eV)+\chi^{\rm R}_{\overline{\sigma},\overline{\sigma}',\overline{\sigma}',\overline{\sigma}}(x,\omega-\omega')\big]n^{}_{\rm F}(\omega+eV)\big[1-n^{}_{\rm F}(\omega')\big]\Big\}\nonumber\\&+e\,\pi  U^{2}\sum_{\sigma}\int\limits_{-\infty}^{\infty}d\omega d\omega'\,\rho^{}_{{\rm L},\overline\sigma}(x,\omega+eV)\rho^{}_{{\rm R},\sigma}(x,\omega')\rho^{\rm II}_{{\rm L},\sigma,\overline\sigma}(x,\omega+\omega'+eV)\nonumber\\&\times\Big\{n^{}_{\rm F}(\omega+eV)n^{}_{\rm F}(\omega')\big[1-n^{}_{\rm F}(\omega+\omega'+eV)\big]-\big[1-n^{}_{\rm F}(\omega+eV)\big]\big[1-n^{}_{\rm F}(\omega')\big]n^{}_{\rm F}(\omega+\omega'+eV)\Big\}\nonumber\\&-e\,\pi  U^{2}\sum_{\sigma}\int\limits_{-\infty}^{\infty}d\omega d\omega'\,\rho^{}_{{\rm L},\overline\sigma}(x,\omega+eV)\rho^{}_{{\rm R},\sigma}(x,\omega')\rho^{\rm II}_{{\rm R},\sigma,\overline\sigma}(x,\omega+\omega')\nonumber\\&\times\Big\{n^{}_{\rm F}(\omega+eV)n^{}_{\rm F}(\omega')\big[1-n^{}_{\rm F}(\omega+\omega')\big]-\big[1-n^{}_{\rm F}(\omega+eV)\big]\big[1-n^{}_{\rm F}(\omega')\big]n^{}_{\rm F}(\omega+\omega')\Big\},
\end{align}
where 
$\chi^{ X}_{\sigma_{1},\sigma_{2},\sigma_{3},\sigma_{4}}(x,t)=\big<\Psi^{\dagger}_{X ,\sigma_{1}}(x,t)\Psi^{}_{X,\sigma_{2}}(x,t)\Psi^{\dagger}_{X ,\sigma_{3}}(x,0)\Psi^{}_{X,\sigma_{4}}(x,0)\big>_{H^{}_{X}}$ and $\rho^{\rm II}_{\sigma,\overline\sigma}(x,\omega)=-\frac{1}{\pi}{\rm Im}\, G^{\rm II}_{\sigma,\overline\sigma}(x,\omega)$ is the two-particle density of states, with
\begin{equation}
G^{\rm II}_{\sigma,\overline\sigma}(x,t)=-i\theta(t)\langle\{\Psi^{}_{\sigma}(x,t)\Psi^{}_{\overline\sigma}(x,t),\Psi^{\dagger}_{\overline\sigma}(0)\Psi^{\dagger}_{\sigma}(0)\}\rangle_{H},
\end{equation}
and $\bar\sigma$ is the opposite of $\sigma$.   Note that $\chi$ is not a time-ordered expectation, but is instead, apart from thermal factors, related to a two-particle spectral function. 
For a system with weak or no correlations on one side and a spin-independent featureless density of states near the Fermi energy $\epsilon^{}_{\rm F}$ and neglecting the contributions from the two-particle density of states,
\begin{align}
\label{final I2b}
I_{2}\approx e U^{2}\rho^{}_{{\rm L}}(\epsilon_{\rm F})\rho^{}_{{\rm R}}(\epsilon_{\rm F})\sum_{\sigma,\sigma'}\int\limits_{-\infty}^{\infty}d\omega d\omega'\, &\chi^{\rm R}_{\sigma',\sigma,\sigma,\sigma'}(x,\omega-\omega')\Big\{n^{}_{\rm F}(\omega)\big[1-n^{}_{\rm F}(\omega'+eV)\big]-n^{}_{\rm F}(\omega+eV)\big[1-n^{}_{\rm F}(\omega')\big]\Big\}.
\end{align}
Equation \eqref{final I2b}  can be rewritten using 
\begin{align}
\sum_{\sigma,\sigma'}\chi^{\rm R}_{\sigma',\sigma,\sigma,\sigma'}(x,t)&=2S_{\rm R}(x,t)+\frac{1}{2}\Pi_{\rm R}(x,t)\nonumber\\&=2\langle{\bf s}^{}_{\rm R}(x,t)\cdot{\bf s}^{}_{\rm R}(x,0)\rangle +\frac{1}{2}\langle\hat{n}^{}_{\rm R}(x,t)\hat{n}^{}_{\rm R}(x,0)\rangle, 
\end{align}
where $\langle{\bf s}^{}_{\rm R}(x,t)\cdot{\bf s}^{}_{\rm R}(x,0)\rangle$ and $\langle\hat{n}^{}_{\rm R}(x,t)\hat{n}^{}_{\rm R}(x,0)\rangle$ are the spin-spin (density) and density-density correlation functions respectively, giving 
\begin{align}
\label{approximate I2}
I_{2}= 2e U^{2}\rho^{}_{{\rm L}}(\epsilon_{\rm F})\rho^{}_{{\rm R}}(\epsilon_{\rm F})\int\limits_{-\infty}^{\infty}d\omega d\omega'\, &S^{}_{\rm R}(x,\omega-\omega')\Big\{n^{}_{\rm F}(\omega)\big[1-n^{}_{\rm F}(\omega'+eV)\big]-n^{}_{\rm F}(\omega+eV)\big[1-n^{}_{\rm F}(\omega')\big]\Big\}\nonumber\\+\frac{e U^{2}\rho^{}_{{\rm L}}(\epsilon_{\rm F})\rho^{}_{{\rm R}}(\epsilon_{\rm F})}{2}\int\limits_{-\infty}^{\infty}d\omega d\omega'\, &\Pi^{}_{\rm R}(x,\omega-\omega')\Big\{n^{}_{\rm F}(\omega)\big[1-n^{}_{\rm F}(\omega'+eV)\big]-n^{}_{\rm F}(\omega+eV)\big[1-n^{}_{\rm F}(\omega')\big]\Big\}.
\end{align}
\end{widetext}

For spin systems, such as those of  Refs.\ [\onlinecite{HeinrichScience04,HirjibehedinScience06,BalashovPRL06}], the density fluctuation term can be safely neglected.  This reduces Eq.\ \eqref{approximate I2} to the form of those proprosed in Refs.\ \onlinecite{{RossierPRL09, FranssonNanoLett09}}, which have accurately described the experiments of  Refs.\ \onlinecite{HeinrichScience04,HirjibehedinScience06}.  But in general, Eq.\ \eqref{I2} gives the total correlated transfer contribution to the tunneling current and can be used in a broader range of experiments, which will be the subject of a future publications.

Within the approximations leading to Eq.\ \eqref{approximate I2} and in the low temperature limit, $k^{}_{\rm B}T \ll \epsilon_{\rm F}$, the differential conductance $G=\partial_{V}I$ is
\begin{align}
G(eV)&=4\pi e^{2}|T|^{2}\rho^{}_{{\rm L}}(\epsilon_{\rm F})\rho^{}_{{\rm R}}(\epsilon_{\rm F})+2e^{2}U^{2}\rho^{}_{{\rm L}}(\epsilon_{\rm F})\rho^{}_{{\rm R}}(\epsilon_{\rm F})\nonumber\\ &\times\bigg\{\int d\omega\, S^{}_{\rm R}(x,\omega)\big[n^{}_{\rm F}(\omega+eV)+n^{}_{\rm F}(\omega-eV)\big]\nonumber\\&+\frac{1}{4}\int d\omega\, \Pi^{}_{\rm R}(x,\omega)\big[n^{}_{\rm F}(\omega+eV)+n^{}_{\rm F}(\omega-eV)\big]\bigg\}.
\end{align}
The second derivative of the current gives 
\begin{align}
\partial^{2}_{V}I&=2e^{3}U^{2}\rho^{}_{{\rm L}}(\epsilon_{\rm F})\rho^{}_{{\rm R}}(\epsilon_{\rm F})\Big\{S^{}_{\rm R}(x,eV)-S^{}_{\rm R}(x,-eV)\nonumber\\&+\frac{1}{4}\big[\Pi^{}_{\rm R}(x,eV)-\Pi^{}_{\rm R}(x,-eV)\big]\Big\}.
\end{align}
\section{\label{sec: discussion} Outlook}
As mentioned in the introduction and demonstrated in the body of the article, in principle all tunneling experiments, at some level, probe two-particle properties induced by interactions effects. Although, by their very nature these effects can be small, when compared to the single-particle terms. It takes a well engineered experiment, such as those of Refs.\ [\onlinecite{HeinrichScience04,HirjibehedinScience06,BalashovPRL06}], to separate the contributions.  Another system where such effects have probably been observed, is in transport through a quantum point contact (QPC).  It is believed that interactions are responsible for the ``0.7 conductance anomaly'' of QPCs.\cite{ThomasPRL96} A lot of theoretical work has gone into explaining this effect.  For instance, in Ref.\ [\onlinecite{ MeirPRL02}]  a generalized Anderson model, which effectively contains correlated transfer of electrons, has been applied with some success to this problem.
In the previous sections we have calculated the contribution from interactions to the many-body tunneling current within an ``STM-like'' geometry.  A natural extension would be to apply the ideas developed here to a quantum-dot geometry or QPC.   Such a derivation would lead to a  generalization of the well-known Meir-Wingreen expression.\cite{MeirPRL92}   

Furthermore, one could use the results presented here to design experiments to probe the two-particle properties of other important and interesting systems.  For example, how these properties change as one approaches the superconducting instability.  This could give information about the pairing mechanism of high-temperature superconductors.       
\begin{acknowledgments}
The author would like to thank Michael Geller, Hartmut Hafermann, Mikhail Katsnelson, and Alexander Licthenstein for many useful conversations.   This work has been supported by the German Research Council (DFG) under SFB 668 and the Louisiana Board of Regents.   
\end{acknowledgments}
\appendix
\section{Details of the derivation of the tunneling Hamiltonian}
\label{Details of the derivation of the tunneling Hamiltonian}
Here we give the details of the derivation of Eq.\ \eqref{tunneling Hamiltonian}.  The kinetic energy term was done in the body of the paper, here we will deal with the interaction term.   Expressing the interaction term of \eqref{exact Hamiltonian} in terms of the projected operators, Eqs.\ \eqref{approximate projected operators} and keeping only terms to leading order in the tunneling amplitude gives
\begin{widetext}
\begin{align}
&\frac{1}{2}\sum_{\sigma,\sigma'}\int dx dx'\,\Psi^{\dagger}_{\sigma}(x)\Psi^{\dagger}_{\sigma'}(x')U(x,x')\Psi^{}_{\sigma'}(x')\Psi^{}_{\sigma}(x)=\nonumber\\& \underbrace{\frac{1}{2}\sum_{\sigma,\sigma'}\int dx dx'\,\Psi^{\dagger}_{{\rm L},\sigma}(x)\Psi^{\dagger}_{{\rm L},\sigma'}(x')U(x,x')\Psi^{}_{{\rm L},\sigma'}(x')\Psi^{}_{{\rm L},\sigma}(x)}_{\text{Left interaction}}+\underbrace{\frac{1}{2}\sum_{\sigma,\sigma'}\int dx dx'\,\Psi^{\dagger}_{{\rm R},\sigma}(x)\Psi^{\dagger}_{{\rm R},\sigma'}(x')U(x,x')\Psi^{}_{{\rm R},\sigma'}(x')\Psi^{}_{{\rm R},\sigma}(x)}_{\text{Right interaction}}\nonumber\\+&\underbrace{\sum_{\sigma,\sigma'}\int dx dx'\,\Psi^{\dagger}_{{\rm L},\sigma}(x)\Psi^{\dagger}_{{\rm R},\sigma'}(x')U(x,x')\Psi^{}_{{\rm R},\sigma'}(x')\Psi^{}_{{\rm L},\sigma}(x)}_{\text{Direct interaction}}+\underbrace{\sum_{\sigma,\sigma'}\int dx dx'\,\Psi^{\dagger}_{{\rm L},\sigma}(x)\Psi^{\dagger}_{{\rm R},\sigma'}(x')U(x,x')\Psi^{}_{{\rm L},\sigma'}(x')\Psi^{}_{{\rm R},\sigma}(x)}_{\text{Exchange interaction}}\nonumber\\+&\underbrace{\sum_{\sigma}\int dx dx'\,U(x,x')\Big\{\Psi^{\dagger}_{{\rm L},\sigma}(x)\Psi^{}_{{\rm R},\sigma}(x)\big[\hat{n}_{\rm L}(x')+\hat{n}_{\rm R}(x')\big]+{\rm H.c.}\Big\}}_{\text{Tunneling}}+O(T^{2}_{k,k'}),
\end{align}
\end{widetext}
where $\hat{n}(x)=\sum_{\sigma}\Psi^{\dagger}_{\sigma}(x)\Psi_{\sigma}(x)$.  The  exchange and direct terms don't contribute to tunneling current, i.e.  they both commute with the number operator of each side, see Sec. \ref{sec:current}, and thus will be neglected. The full effective Hamiltonian then reduces to that given by   Eq.\ \eqref{tunneling Hamiltonian}.

\section{Details of the derivation of the tunneling current}
\label{tunneling current appendix}

Here the details are given for the evaluation of the tunneling current, Eq. \eqref{LR current}.  Although the evaluation is quite lengthy and tedious, a lot of details are included for completeness.

Defining $\hat{A}_{\sigma}(x)=T(x)\Psi^{\dagger}_{{\rm L},\sigma}(x)\Psi^{}_{{\rm R},\sigma}(x)$ and $\hat{B}_{\sigma}(x,x')=U(x,x')\Psi^{\dagger}_{{\rm L},\sigma}(x)\Psi^{}_{{\rm R},\sigma}(x)\big[\hat{n}_{\rm L}(x')+\hat{n}_{\rm L}(x')\big]$ for notational convenance. Note that
\begin{align*}
\langle\hat{A}_{\sigma}(x)\rangle^{}_{H_{0}}=\langle\hat{B}_{\sigma}(x,x')\rangle^{}_{H_{0}}&=0\nonumber\\ 
\langle\hat{A}_{\sigma}(x)\hat{A}_{\sigma'}(x')\rangle^{}_{H_{0}}=\langle\hat{B}_{\sigma}(x,x')\hat{B}_{\sigma'}(x'',x''')\rangle^{}_{H_{0}}&=0
\end{align*}
and
\begin{align*}
\langle\hat{A}_{\sigma}(x)\hat{B}_{\sigma'}(x',x'')\rangle^{}_{H_{0}}&=0.
\end{align*}
  Evaluating the commutator in Eq.~\eqref{LR current}, the total current can be written as $I=I_{1}+I_{2}+I_{3}$, where
 \begin{align}
 \label{appendix I1}
 I_{1}=2e\int \limits_{-\infty}^{t}dt'\sum_{\sigma,\sigma'}\int dx dx'\,& {\rm Re}\Big[\langle\hat{A}^{\dagger}_{\sigma'}(x',t')\hat{A}^{}_{\sigma}(x,t)\rangle\nonumber\\&-\langle\hat{A}^{}_{\sigma'}(x',t')\hat{A}^{\dagger}_{\sigma}(x,t)\rangle\Big],
 \end{align}
 \begin{align}
 \label{appendix I2}
 I_{2}=2e\int \limits_{-\infty}^{t}dt'&\sum_{\sigma,\sigma'}\int dx_{1} dx_{2}dx_{3}dx_{4}\,\nonumber\\&\times{\rm Re}\Big[\langle\hat{B}^{\dagger}_{\sigma'}(x_{1},x_{2},t')\hat{B}^{}_{\sigma}(x_{3},x_{4},t)\rangle\nonumber\\&-\langle\hat{B}^{}_{\sigma'}(x_{1},x_{2},t')\hat{B}^{\dagger}_{\sigma}(x_{3},x_{4},t)\rangle\Big],
 \end{align}
 and
 \begin{align}
 \label{appendix I3}
 &I_{3}=2e\int \limits_{-\infty}^{t}dt'\sum_{\sigma,\sigma'}\int dx' dx_{1}dx_{2}\,\nonumber\\&{\rm Re}\Big[\langle\hat{A}^{\dagger}_{\sigma'}(x',t')\hat{B}^{}_{\sigma}(x_{1},x_{2},t)\rangle-\langle{A}^{}_{\sigma'}(x',t')\hat{B}^{\dagger}_{\sigma}(x_{1},x_{2},t)\rangle\nonumber\\&+\langle \hat{B}^{\dagger}_{\sigma'}(x_{1},x_{2},t')\hat{A}_{\sigma}(x',t)\rangle-\langle \hat{B}^{}_{\sigma'}(x_{1},x_{2},t')\hat{A}^{\dagger}_{\sigma}(x',t)\rangle\Big].
 \end{align}
 The first two contributions will be evaluated in the following sections, while the third term, $I_{3}$, will be neglected. As it contains two-particle correlations with three equal times. It would be expected that these correlations are subdominant by this restriction of phase-space.
\subsection{$I_{1}$}
To evaluate the expectation values in Eq.~\eqref{appendix I1} it is useful to introduce  the Keldysh contour-ordered single-particle Green's function\cite{Haugbook}
\begin{equation}
G^{X}_{\sigma}(x,t,t')=-i\big<T^{}_{C}\Psi^{}_{X,\sigma}(x,t)\Psi^{\dagger}_{X,\sigma}(x,t')\big>_{H^{}_{X}},
\end{equation} 
where $X\in \{{\rm L},{\rm R}\}$. The quantities of interest for this work are the ``lesser'' $(<)$ and ``greater'' $(>)$ functions  given by 
\begin{align}
\label{lesser GF}
\big[G^{X}_{\sigma}(x,t,t')\big]^{<}&=i\big<\Psi^{\dagger}_{X,\sigma}(x,t')\Psi^{}_{X,\sigma}(x,t)\big>_{H^{}_{X}}\nonumber\\&=i\int d\omega\, e^{-i\omega (t-t')}n^{}_{\rm F}(\omega)\rho^{}_{X,\sigma}(x,\omega)
\end{align}
and
\begin{align}
\label{greater GF}
 &\big[G^{X}_{\sigma}(x,t,t')\big]^{>}=-i\big<\Psi^{}_{X,\sigma}(x,t)\Psi^{\dagger}_{X,\sigma}(x,t')\big>_{H^{}_{X}}\nonumber\\&=-i\int d\omega\, e^{-i\omega(t-t')}\big[1-n^{}_{\rm F}(\omega)\big]\rho^{}_{X,\sigma}(x,\omega),
\end{align}
where $n^{}_{\rm F}(\omega)=(e^{\beta\omega}+1)^{-1}$ is the Fermi distribution with inverse temperature $\beta=(k_{\rm B}T)^{-1}$ and $\rho^{\rm I}_{\sigma}(x,\omega)$ is the single-particle local density of states, obtained from the imaginary part of the retarded Green's function,
\begin{equation}
\rho^{}_{\sigma}(x,\omega)=-\frac{1}{\pi}{\rm Im}\, G^{\rm ret}_{\sigma}(x,\omega).
\end{equation}

In the limit were the tunneling is spatially localized about a single point, as it for an STM, where the wave functions of the tip are exponentially  localized on the atomic scale about the location of the tip, the spatial integrals in \eqref{appendix I1} can be approximated by the value of the functions centered at the tunneling point.   This is equivalent to the, common, assumption that the tunneling matrix elements $T_{{\bf k},{\bf k}'}=\langle \psi_{{\rm L},{\bf k}}|T|\psi_{{\rm R},{\bf k}'}\rangle$ have no momentum dependance.  The needed correlation functions from \eqref{appendix I1} are then (assuming spin is conserved during tunneling)
\begin{align}
&\langle\hat{A}^{\dagger}_{\sigma}(x,t')\hat{A}^{}_{\sigma}(x,t)\rangle_{H_{0}}=\nonumber\\&|T|^{2}\langle\Psi^{}_{{\rm L},\sigma }(x,t')\Psi^{\dagger}_{{\rm L},\sigma }(x,t)\rangle_{H_{\rm L}}\langle\Psi^{\dagger}_{{\rm R},\sigma }(x,t')\Psi^{}_{{\rm R},\sigma }(x,t)\rangle_{H_{\rm R}}
\end{align}
and
\begin{align}
&\langle\hat{A}^{}_{\sigma}(x,t')\hat{A}^{\dagger}_{\sigma}(x,t)\rangle_{H_{0}}=\nonumber\\&|T|^{2}\langle\Psi^{\dagger}_{{\rm L},\sigma}(x,t')\Psi^{}_{{\rm L},\sigma}(x,t)\rangle_{H_{\rm L}}\langle\Psi^{}_{{\rm R},\sigma}(x,t')\Psi^{\dagger}_{{\rm R},\sigma}(x,t)\rangle_{H_{\rm R}}.
\end{align}
By using  Eqs.~\eqref{lesser GF} and \eqref{greater GF},  the well-known result relating the tunneling current to the  single-particle local density of states is recovered,
\begin{align}
I_{1}&=2\pi e|T|^{2}\sum_{\sigma}\int d\omega\,\rho^{}_{{\rm L},\sigma}(x,\omega+eV)\rho^{}_{{\rm R},\sigma}(x,\omega)\nonumber\\ &\times\big[n^{}_{\rm F}(\omega)-n^{}_{\rm F}(\omega+eV)\big].
\end{align}
\subsection{$I_{2}$}
The remaining contribution to the tunneling current, $I_{2}$, contains many  high order correlations functions, including three particle propagators.  The evaluations of which becomes quite tedious and lengthy.  But as this is directly related to this work, we will cover these in some detail.   

From Eq.\ \eqref{appendix I2}, the needed correlations for $I_{2}$ are 
\begin{widetext}
\begin{align}
\langle\hat{B}^{\dagger}_{\sigma}(x_{1},x_{2},t')\hat{B}^{}_{\sigma}(x_{3},x_{4},t)\rangle^{}_{H_{0}}&=\nonumber\\ &\hspace{-1.0cm}U(x_{1},x_{2})U(x_{3},x_{4})\Big[\langle\hat{n}^{}_{\rm L}(x_{2},t')\Psi^{}_{{\rm L},\sigma}(x_{1},t')\Psi^{\dagger}_{{\rm L},\sigma}(x_{3},t)\hat{n}^{}_{\rm L}(x_{4},t)\rangle^{}_{H^{}_{\rm L}}\langle\Psi^{\dagger}_{{\rm R},\sigma}(x_{1},t')\Psi^{}_{{\rm R},\sigma}(x_{3},t)\rangle^{}_{H_{\rm R}}\nonumber\\&\hspace{-1.0cm}+\langle\Psi^{}_{{\rm L},\sigma}(x_{1},t')\Psi^{\dagger}_{{\rm L},\sigma}(x_{3},t)\rangle^{}_{H^{}_{\rm L}}\langle\hat{n}^{}_{\rm R}(x_{2},t')\Psi^{\dagger}_{{\rm R},\sigma}(x_{1},t')\Psi^{}_{{\rm R},\sigma}(x_{3},t)\hat{n}^{}_{\rm R}(x_{4},t)\rangle^{}_{H_{\rm R}}\nonumber\\&\hspace{-1.0cm}+\langle\hat{n}^{}_{\rm L}(x_{2},t')\Psi^{}_{{\rm L},\sigma}(x_{1},t')\Psi^{\dagger}_{{\rm L},\sigma}(x_{3},t)\rangle^{}_{H_{\rm L}}\langle\Psi^{\dagger}_{{\rm R},\sigma}(x_{1},t')\Psi^{}_{{\rm R},\sigma}(x_{3},t)\hat{n}^{}_{\rm R}(x_{4},t)\rangle^{}_{H_{\rm R}}\nonumber\\&\hspace{-1.0cm}+\langle\Psi^{}_{{\rm L},\sigma}(x_{1},t')\Psi^{\dagger}_{{\rm L},\sigma}(x_{3},t)\hat{n}^{}_{\rm L}(x_{4},t)\rangle^{}_{H_{\rm L}}\langle\hat{n}^{}_{\rm R}(x_{2},t')\Psi^{\dagger}_{{\rm R},\sigma}(x_{1},t')\Psi^{}_{{\rm R},\sigma}(x_{3},t)\rangle^{}_{H_{\rm R}}\Big],
\end{align}
and
\begin{align}
\langle\hat{B}^{}_{\sigma}(x_{1},x_{2},t')\hat{B}^{\dagger}_{\sigma}(x_{3},x_{4},t)\rangle^{}_{H_{0}}=\nonumber\\ &\hspace{-1.0cm}U(x_{1},x_{2})U(x_{3},x_{4})\Big[\langle\Psi^{\dagger}_{{\rm L},\sigma}(x_{1},t')\hat{n}^{}_{\rm L}(x_{2},t')\hat{n}^{}_{\rm L}(x_{4},t)\Psi^{}_{{\rm L},\sigma}(x_{3},t)\rangle^{}_{H_{\rm L}}\langle\Psi^{}_{{\rm R},\sigma}(x_{1},t')\Psi^{\dagger}_{{\rm R},\sigma}(x_{3},t)\rangle^{}_{H_{\rm R}}\nonumber\\&\hspace{-1.0cm}+\langle\Psi^{\dagger}_{{\rm L},\sigma}(x_{1},t')\Psi^{}_{{\rm L},\sigma}(x_{3},t)\rangle^{}_{H_{\rm L}}\langle\Psi^{}_{{\rm R},\sigma}(x_{1},t')\hat{n}^{}_{\rm R}(x_{2},t')\hat{n}^{}_{\rm R}(x_{4},t)\Psi^{\dagger}_{{\rm R},\sigma}(x_{3},t)\rangle^{}_{H_{\rm R}}\nonumber\\&\hspace{-1.0cm}+\langle\Psi^{\dagger}_{{\rm L},\sigma}(x_{1},t')\hat{n}^{}_{\rm L}(x_{2},t')\Psi^{}_{{\rm L},\sigma}(x_{3},t)\rangle_{H^{}_{\rm L}}\langle\Psi^{}_{{\rm R},\sigma}(x_{1},t')\hat{n}^{}_{\rm R}(x_{4},t)\Psi^{\dagger}_{{\rm R},\sigma}(x_{3},t)\rangle_{H^{}_{\rm R}}\nonumber\\&\hspace{-1.0cm}+\langle\Psi^{\dagger}_{{\rm L},\sigma}(x_{1},t')\hat{n}^{}_{\rm L}(x_{4},t)\Psi^{}_{{\rm L},\sigma}(x_{3},t)\rangle_{H^{}_{\rm L}}\langle\Psi^{}_{{\rm R},\sigma}(x_{1},t')\hat{n}^{}_{\rm R}(x_{2},t')\Psi^{\dagger}_{{\rm R},\sigma}(x_{3},t)\rangle_{H^{}_{\rm R}}\Big]. 
\end{align}
Neglecting  all two-particle correlations that have three equal times, as these are expected to be sub-dominate from the reduction of phase space,  leads to
\begin{align}
\langle\hat{B}^{\dagger}_{\sigma}(x_{1},x_{2},t')\hat{B}^{}_{\sigma}(x_{3},x_{4},t)\rangle^{}_{H_{0}}&\approx\nonumber\\ &\hspace{-1.0cm}U(x_{1},x_{2})U(x_{3},x_{4})\Big[\langle\hat{n}^{}_{\rm L}(x_{2},t')\Psi^{}_{{\rm L},\sigma}(x_{1},t')\Psi^{\dagger}_{{\rm L},\sigma}(x_{3},t)\hat{n}^{}_{\rm L}(x_{4},t)\rangle^{}_{H^{}_{\rm L}}\langle\Psi^{\dagger}_{{\rm R},\sigma}(x_{1},t')\Psi^{}_{{\rm R},\sigma}(x_{3},t)\rangle^{}_{H_{\rm R}}\nonumber\\&\hspace{-1.0cm}+\langle\Psi^{}_{{\rm L},\sigma}(x_{1},t')\Psi^{\dagger}_{{\rm L},\sigma}(x_{3},t)\rangle^{}_{H^{}_{\rm L}}\langle\hat{n}^{}_{\rm R}(x_{2},t')\Psi^{\dagger}_{{\rm R},\sigma}(x_{1},t')\Psi^{}_{{\rm R},\sigma}(x_{3},t)\hat{n}^{}_{\rm R}(x_{4},t)\rangle^{}_{H_{\rm R}}\Big]
\end{align}
and
\begin{align}
\langle\hat{B}^{}_{\sigma}(x_{1},x_{2},t')\hat{B}^{\dagger}_{\sigma}(x_{3},x_{4},t)\rangle^{}_{H_{0}}\approx\nonumber\\ &\hspace{-1.0cm}U(x_{1},x_{2})U(x_{3},x_{4})\Big[\langle\Psi^{\dagger}_{{\rm L},\sigma}(x_{1},t')\hat{n}^{}_{\rm L}(x_{2},t')\hat{n}^{}_{\rm L}(x_{4},t)\Psi^{}_{{\rm L},\sigma}(x_{3},t)\rangle^{}_{H_{\rm L}}\langle\Psi^{}_{{\rm R},\sigma}(x_{1},t')\Psi^{\dagger}_{{\rm R},\sigma}(x_{3},t)\rangle^{}_{H_{\rm R}}\nonumber\\&\hspace{-1.0cm}+\langle\Psi^{\dagger}_{{\rm L},\sigma}(x_{1},t')\Psi^{}_{{\rm L},\sigma}(x_{3},t)\rangle^{}_{H_{\rm L}}\langle\Psi^{}_{{\rm R},\sigma}(x_{1},t')\hat{n}^{}_{\rm R}(x_{2},t')\hat{n}^{}_{\rm R}(x_{4},t)\Psi^{\dagger}_{{\rm R},\sigma}(x_{3},t)\rangle^{}_{H_{\rm R}}\Big].
\end{align}
\end{widetext}

To evaluate the remaining  three-particle correlations it will be useful to introduce the following contour-order quantities 
\begin{align}
\label{CL}
C^{\rm L}_{\sigma}(t',t)&=\nonumber\\&\langle T_{C}\hat{n}^{}_{\rm L}(x_{1},t')\Psi^{}_{{\rm L},\sigma}(x_{2},t')\Psi^{\dagger}_{{\rm L},\sigma}(x_{3},t)\hat{n}^{}_{\rm L}(x_{4},t)\rangle^{}_{H^{}_{\rm L}},
\end{align}
and
\begin{align}
\label{CR}
C^{\rm R}_{\sigma}(t',t)&=\nonumber\\&\langle T_{C}\hat{n}^{}_{\rm L}(x_{1},t')\Psi^{\dagger}_{{\rm R},\sigma}(x_{2},t')\Psi^{}_{{\rm R},\sigma}(x_{3},t)\hat{n}^{}_{\rm R}(x_{4},t)\rangle^{}_{H^{}_{\rm R}}.
\end{align}
Such that
\begin{align}
\big[C^{\rm L}_{\sigma}(t',t)\big]^{>}&=\nonumber\\&\!\langle \hat{n}^{}_{\rm L}(x_{1},t')\Psi^{}_{{\rm L},\sigma}(x_{2},t')\Psi^{\dagger}_{{\rm L},\sigma}(x_{3},t)\hat{n}^{}_{\rm L}(x_{4},t)\rangle_{H^{}_{\rm L}},
\end{align}
\begin{align}
\big[C^{\rm L}_{\sigma}(t',t)&\big]^{<}=\nonumber\\&-\langle \Psi^{\dagger}_{{\rm L},\sigma}(x_{3},t)\hat{n}^{}_{\rm L}(x_{4},t)\hat{n}^{}_{\rm L}(x_{1},t')\Psi^{}_{{\rm L},\sigma}(x_{2},t')\rangle_{H^{}_{\rm L}},
\end{align}
and
\begin{align}
\big[C^{\rm R}_{\sigma}(t',t)\big]^{>}&=\nonumber\\&\hspace{-0.2cm}\langle \hat{n}^{}_{\rm R}(x_{1},t')\Psi^{\dagger}_{{\rm R},\sigma}(x_{2},t')\Psi^{}_{{\rm R},\sigma}(x_{3},t)\hat{n}^{}_{\rm R}(x_{4},t)\rangle_{H^{}_{\rm R}},
\end{align}
\begin{align}
\big[C^{\rm R}_{\sigma}(t',t)&\big]^{<}=\nonumber\\&\hspace{-0.23cm}-\langle \Psi^{}_{{\rm R},\sigma}(x_{3},t)\hat{n}^{}_{\rm R}(x_{4},t)\hat{n}^{}_{\rm R}(x_{1},t')\Psi^{\dagger}_{{\rm R},\sigma}(x_{2},t')\rangle_{H^{}_{\rm R}}.
\end{align}
The three-particle correlation functions \eqref{CL} and \eqref{CR} can be expressed as, neglecting two-particle correlations with three equal times and the three-particle vertex,

\begin{widetext}
 \begin{align}
C^{\rm L}_{\sigma}(t',t)&\approx\sum_{\sigma'}\langle T^{}_{C}\Psi^{\dagger}_{{\rm L},\sigma'}(x_{1},t')\Psi_{{\rm L},\sigma'}(x_{4},t)\rangle\langle T^{}_{C}\Psi^{}_{{\rm L},\sigma'}(x_{1},t')\Psi^{}_{{\rm L},\sigma}(x_{2},t')\Psi^{\dagger}_{{\rm L},\sigma}(x_{3},t)\Psi^{\dagger}_{{\rm L},\sigma'}(x_{4},t)\rangle\nonumber\\&-\sum_{\sigma'}\langle T^{}_{C}\Psi^{}_{{\rm L},\sigma}(x_{1},t')\Psi^{\dagger}_{{\rm L},\sigma}(x_{3},t)\rangle\langle T^{}_{C}\Psi^{\dagger}_{{\rm L},\sigma}(x_{1},t')\Psi^{}_{{\rm L},\sigma}(x_{2},t')\Psi^{\dagger}_{{\rm L},\sigma'}(x_{4},t)\Psi^{}_{{\rm L},\sigma'}(x_{4},t)\rangle\nonumber\\&+\sum_{\sigma'}\langle T^{}_{C}\Psi^{}_{{\rm L},\sigma'}(x_{1},t')\Psi^{\dagger}_{{\rm L},\sigma'}(x_{4},t)\rangle\langle T^{}_{C}\Psi^{\dagger}_{{\rm L},\sigma'}(x_{1},t')\Psi^{}_{{\rm L},\sigma}(x_{2},t')\Psi^{\dagger}_{{\rm L},\sigma}(x_{3},t)\Psi^{}_{{\rm L},\sigma'}(x_{4},t)\rangle\nonumber\\&-\sum_{\sigma'}\langle T^{}_{C}\Psi^{}_{{\rm L},\sigma}(x_{2},t')\Psi^{\dagger}_{{\rm L},\sigma}(x_{4},t)\rangle\langle T^{}_{C}\Psi^{\dagger}_{{\rm L},\sigma'}(x_{1},t')\Psi^{}_{{\rm L},\sigma'}(x_{1},t')\Psi^{\dagger}_{{\rm L},\sigma}(x_{3},t)\Psi^{}_{{\rm L},\sigma}(x_{4},t)\rangle\nonumber\\&+\langle T^{}_{C}\Psi^{}_{{\rm L},\sigma}(x_{2},t')\Psi^{\dagger}_{{\rm L},\sigma}(x_{3},t)\rangle\langle T^{}_{C}\hat{n}^{}_{\rm L}(x_{1},t')\hat{n}^{}_{\rm L}(x_{4},t)\rangle.  
\end{align}
A similar expression exists for $C^{\rm R}_{\sigma}(t',t)$, Eq.~\eqref{CR}. 

Again assuming we are in the point contact or localized tunneling regime, where the wave functions of one side, say L, are spatially localized about the tunneling point, $x$, the dominate contribution  from the spatial integrals of \eqref{appendix I2} can be approximated by the value at $x$.  This is equivalent  to neglecting the momentum dependance of the matrix elements of the interaction $U_{{\bf k},{\bf k}'}=\langle \psi_{{\rm L},{\bf k}}|U|\psi_{{\rm R},{\bf k}}\rangle$, which for a good metal is approximately true; especially for the small energy region about the Fermi energy that one is typically interested in for tunneling experiments.   Also for a contour-ordered quantity such as $X(t,t')=Y(t,t')Z(t,t')$, the greater and lesser functions are simply giving by $\big[X(t,t')]^{\gtrless}=\big[Y(t,t')\big]^{\gtrless}\big[Z(t,t')\big]^{\gtrless}$. Along with defining 
\begin{equation}
\chi^{X}_{\sigma_{1},\sigma_{2},\sigma_{3},\sigma_{4}}(x,t-t')=\langle \Psi^{\dagger}_{X,\sigma_{1}}(x,t)\Psi^{}_{X,\sigma_{2}}(x,t)\Psi^{\dagger}_{X,\sigma_{3}}(x,t')\Psi^{}_{X,\sigma_{4}}(x,t')\rangle=\int \frac{d\omega}{2\pi}\, \chi^{X}_{\sigma_{1},\sigma_{2},\sigma_{3},\sigma_{4}}(x,\omega)e^{-i\omega (t-t')},
\end{equation}
and the two-particle density of states $\rho^{\rm II}_{\sigma,\overline\sigma}(x,\omega)=-\frac{1}{\pi}{\rm Im}\, G^{\rm II}_{\sigma,\overline\sigma}(x,\omega)$, where
\begin{equation}
G^{\rm II}_{\sigma,\overline\sigma}(x,t)=-i\theta(t)\langle\{\Psi^{}_{\sigma}(x,t)\Psi^{}_{\overline\sigma}(x,t),\Psi^{\dagger}_{\overline\sigma}(0)\Psi^{\dagger}_{\sigma}(0)\}\rangle_{H},
\end{equation}
Eq.~\eqref{appendix I2} can finally be evaluated and is giving by
\begin{align}
\label{appendix final I2a}
&I_{2}=\nonumber\\&e U^{2}\sum_{\sigma,\sigma'}\int\limits_{-\infty}^{\infty}d\omega d\omega'\, \rho^{}_{{\rm L},\sigma}(x,\omega+eV)\rho^{}_{{\rm R},\sigma'}(x,\omega')\Big\{\big[\chi^{\rm L}_{\overline{\sigma}',\overline{\sigma},\overline{\sigma},\overline{\sigma}'}(x,\omega'-\omega-eV)+\chi^{\rm R}_{\overline{\sigma}',\overline{\sigma},\overline{\sigma},\overline{\sigma}'}(x,\omega'-\omega)\big]n^{}_{\rm F}(\omega')\big[1-n^{}_{\rm F}(\omega+eV)\big]\nonumber\\&-\big[\chi^{\rm L}_{\overline{\sigma},\overline{\sigma}',\overline{\sigma}',\overline{\sigma}}(x,\omega-\omega'+eV)+\chi^{\rm R}_{\overline{\sigma},\overline{\sigma}',\overline{\sigma}',\overline{\sigma}}(x,\omega-\omega')\big]n^{}_{\rm F}(\omega+eV)\big[1-n^{}_{\rm F}(\omega')\big]\Big\}\nonumber\\&+e\,\pi  U^{2}\sum_{\sigma}\int\limits_{-\infty}^{\infty}d\omega d\omega'\,\rho^{}_{{\rm L},\overline\sigma}(x,\omega+eV)\rho^{}_{{\rm R},\sigma}(x,\omega')\rho^{\rm II}_{{\rm L},\sigma,\overline\sigma}(x,\omega+\omega'+eV)\nonumber\\&\times\Big\{n^{}_{\rm F}(\omega+eV)n^{}_{\rm F}(\omega')\big[1-n^{}_{\rm F}(\omega+\omega'+eV)\big]-\big[1-n^{}_{\rm F}(\omega+eV)\big]\big[1-n^{}_{\rm F}(\omega')\big]n^{}_{\rm F}(\omega+\omega'+eV)\Big\}\nonumber\\&-e\,\pi  U^{2}\sum_{\sigma}\int\limits_{-\infty}^{\infty}d\omega d\omega'\,\rho^{}_{{\rm L},\overline\sigma}(x,\omega+eV)\rho^{}_{{\rm R},\sigma}(x,\omega')\rho^{\rm II}_{{\rm R},\sigma,\overline\sigma}(x,\omega+\omega')\nonumber\\&\times\Big\{n^{}_{\rm F}(\omega+eV)n^{}_{\rm F}(\omega')\big[1-n^{}_{\rm F}(\omega+\omega')\big]-\big[1-n^{}_{\rm F}(\omega+eV)\big]\big[1-n^{}_{\rm F}(\omega')\big]n^{}_{\rm F}(\omega+\omega')\Big\},
\end{align}
\end{widetext}
which is Eq.\ \eqref{I2}.
\bibliography{/Users/kpatton/Bibliographies/Master}

\begin{thebibliography}{10}%
\makeatletter
\providecommand \@ifxundefined [1]{%
 \ifx #1\undefined \expandafter \@firstoftwo
 \else \expandafter \@secondoftwo
\fi
}%
\providecommand \@ifnum [1]{%
 \ifnum #1\expandafter \@firstoftwo
 \else \expandafter \@secondoftwo
\fi
}%
\providecommand \enquote [1]{``#1''}%
\providecommand \bibnamefont  [1]{#1}%
\providecommand \bibfnamefont [1]{#1}%
\providecommand \citenamefont [1]{#1}%
\providecommand\href[0]{\@sanitize\@href}%
\providecommand\@href[1]{\endgroup\@@startlink{#1}\endgroup\@@href}%
\providecommand\@@href[1]{#1\@@endlink}%
\providecommand \@sanitize [0]{\begingroup\catcode`\&12\catcode`\#12\relax}%
\@ifxundefined \pdfoutput {\@firstoftwo}{%
 \@ifnum{\z@=\pdfoutput}{\@firstoftwo}{\@secondoftwo}%
}{%
 \providecommand\@@startlink[1]{\leavevmode\special{html:<a href="#1">}}%
 \providecommand\@@endlink[0]{\special{html:</a>}}%
}{%
 \providecommand\@@startlink[1]{%
  \leavevmode
  \pdfstartlink
   attr{/Border[0 0 1 ]/H/I/C[0 1 1]}%
   user{/Subtype/Link/A<</Type/Action/S/URI/URI(#1)>>}%
  \relax
 }%
 \providecommand\@@endlink[0]{\pdfendlink}%
}%
\providecommand \url  [0]{\begingroup\@sanitize \@url }%
\providecommand \@url [1]{\endgroup\@href {#1}{\urlprefix}}%
\providecommand \urlprefix [0]{URL }%
\providecommand \Eprint[0]{\href }%
\@ifxundefined \urlstyle {%
  \providecommand \doi [1]{doi:\discretionary{}{}{}#1}%
}{%
  \providecommand \doi [0]{doi:\discretionary{}{}{}\begingroup
  \urlstyle{rm}\Url }%
}%
\providecommand \doibase [0]{http://dx.doi.org/}%
\providecommand \Doi[1]{\href{\doibase#1}}%
\providecommand \bibAnnote [3]{%
  \BibitemShut{#1}%
  \begin{quotation}\noindent
    \textsc{Key:}\ #2\\\textsc{Annotation:}\ #3%
  \end{quotation}%
}%
\providecommand \bibAnnoteFile [2]{%
  \IfFileExists{#2}{\bibAnnote {#1} {#2} {\input{#2}}}{}%
}%
\providecommand \typeout [0]{\immediate \write \m@ne }%
\providecommand \selectlanguage [0]{\@gobble}%
\providecommand \bibinfo [0]{\@secondoftwo}%
\providecommand \bibfield [0]{\@secondoftwo}%
\providecommand \translation [1]{[#1]}%
\providecommand \BibitemOpen[0]{}%
\providecommand \bibitemStop [0]{}%
\providecommand \bibitemNoStop [0]{.\EOS\space}%
\providecommand \EOS [0]{\spacefactor3000\relax}%
\providecommand \BibitemShut [1]{\csname bibitem#1\endcsname}%
\bibitem{GiaeverPRL60}%
  \BibitemOpen
  \bibfield{author}{%
  \bibinfo {author} {\bibfnamefont{I.}~\bibnamefont{Gi$\ae$ver}},\ }%
  \bibfield{journal}{%
  \bibinfo {journal} {Phys. Rev. Lett.}\ }%
  \textbf{\bibinfo {volume} {5}},\ \bibinfo {pages} {464} (\bibinfo {year}
  {1960})%
  \bibAnnoteFile{NoStop}{GiaeverPRL60}%
\bibitem{BardeenPRL61}%
  \BibitemOpen
  \bibfield{author}{%
  \bibinfo {author} {\bibfnamefont{J.}~\bibnamefont{Bardeen}},\ }%
  \bibfield{journal}{%
  \bibinfo {journal} {Phys. Rev. Lett.}\ }%
  \textbf{\bibinfo {volume} {6}},\ \bibinfo {pages} {57} (\bibinfo {year}
  {1961})%
  \bibAnnoteFile{NoStop}{BardeenPRL61}%
\bibitem{CohenPRL62}%
  \BibitemOpen
  \bibfield{author}{%
  \bibinfo {author} {\bibfnamefont{M.~H.}\ \bibnamefont{Cohen}}, \bibinfo
  {author} {\bibfnamefont{L.~M.}\ \bibnamefont{Falicov}},\ and\ \bibinfo
  {author} {\bibfnamefont{J.~C.}\ \bibnamefont{Phillips}},\ }%
  \bibfield{journal}{%
  \bibinfo {journal} {Phys. Rev. Lett.}\ }%
  \textbf{\bibinfo {volume} {8}},\ \bibinfo {pages} {316} (\bibinfo {year}
  {1962})%
  \bibAnnoteFile{NoStop}{CohenPRL62}%
\bibitem{PrangePR63}%
  \BibitemOpen
  \bibfield{author}{%
  \bibinfo {author} {\bibfnamefont{R.~E.}\ \bibnamefont{Prange}},\ }%
  \bibfield{journal}{%
  \bibinfo {journal} {Phys. Rev.}\ }%
  \textbf{\bibinfo {volume} {131}},\ \bibinfo {pages} {1083} (\bibinfo {year}
  {1963})%
  \bibAnnoteFile{NoStop}{PrangePR63}%
\bibitem{ZawadowskiPR67}%
  \BibitemOpen
  \bibfield{author}{%
  \bibinfo {author} {\bibfnamefont{A.}~\bibnamefont{Zawadowski}},\ }%
  \bibfield{journal}{%
  \bibinfo {journal} {Phys. Rev.}\ }%
  \textbf{\bibinfo {volume} {163}},\ \bibinfo {pages} {341} (\bibinfo {year}
  {1967})%
  \bibAnnoteFile{NoStop}{ZawadowskiPR67}%
\bibitem{AppelbaumPR69}%
  \BibitemOpen
  \bibfield{author}{%
  \bibinfo {author} {\bibfnamefont{J.~A.}\ \bibnamefont{Appelbaum}}\ and\
  \bibinfo {author} {\bibfnamefont{W.~F.}\ \bibnamefont{Brinkman}},\ }%
  \bibfield{journal}{%
  \bibinfo {journal} {Phys. Rev.}\ }%
  \textbf{\bibinfo {volume} {186}},\ \bibinfo {pages} {464} (\bibinfo {year}
  {1969})%
  \bibAnnoteFile{NoStop}{AppelbaumPR69}%
\bibitem{CaroliJPhysC71a}%
  \BibitemOpen
  \bibfield{author}{%
  \bibinfo {author} {\bibfnamefont{C.}~\bibnamefont{Caroli}}, \bibinfo {author}
  {\bibfnamefont{R.}~\bibnamefont{Combescot}}, \bibinfo {author}
  {\bibfnamefont{P.}~\bibnamefont{Nozi\`eres}},\ and\ \bibinfo {author}
  {\bibfnamefont{D.}~\bibnamefont{Saint-James}},\ }%
  \bibfield{journal}{%
  \bibinfo {journal} {J. Phys. C: Solid State Phys.}\ }%
  \textbf{\bibinfo {volume} {4}},\ \bibinfo {pages} {916} (\bibinfo {year}
  {1971})%
  \bibAnnoteFile{NoStop}{CaroliJPhysC71a}%
\bibitem{FeuchtwantPRB74a}%
  \BibitemOpen
  \bibfield{author}{%
  \bibinfo {author} {\bibfnamefont{T.~E.}\ \bibnamefont{Feuchtwang}},\ }%
  \bibfield{journal}{%
  \bibinfo {journal} {Phys. Rev. B}\ }%
  \textbf{\bibinfo {volume} {10}},\ \bibinfo {pages} {4121} (\bibinfo {year}
  {1974})%
  \bibAnnoteFile{NoStop}{FeuchtwantPRB74a}%
\bibitem{PerssonPRL87}%
  \BibitemOpen
  \bibfield{author}{%
  \bibinfo {author} {\bibfnamefont{B.~N.~J.}\ \bibnamefont{Persson}}\ and\
  \bibinfo {author} {\bibfnamefont{A.}~\bibnamefont{Baratoff}},\ }%
  \bibfield{journal}{%
  \bibinfo {journal} {Phys. Rev. Lett.}\ }%
  \textbf{\bibinfo {volume} {59}},\ \bibinfo {pages} {339} (\bibinfo {year}
  {1987})%
  \bibAnnoteFile{NoStop}{PerssonPRL87}%
\bibitem{HeinrichScience04}%
  \BibitemOpen
  \bibfield{author}{%
  \bibinfo {author} {\bibfnamefont{A.~J.}\ \bibnamefont{Heinrich}}, \bibinfo
  {author} {\bibfnamefont{J.~A.}\ \bibnamefont{Gupta}}, \bibinfo {author}
  {\bibfnamefont{C.~P.}\ \bibnamefont{Lutz}},\ and\ \bibinfo {author}
  {\bibfnamefont{D.~M.}\ \bibnamefont{Eigler}},\ }%
  \bibfield{journal}{%
  \bibinfo {journal} {Science}\ }%
  \textbf{\bibinfo {volume} {306}},\ \bibinfo {pages} {466} (\bibinfo {year}
  {2004})%
  \bibAnnoteFile{NoStop}{HeinrichScience04}%
\bibitem{HirjibehedinScience06}%
  \BibitemOpen
  \bibfield{author}{%
  \bibinfo {author} {\bibfnamefont{C.~F.}\ \bibnamefont{Hirjibehedin}},
  \bibinfo {author} {\bibfnamefont{C.~P.}\ \bibnamefont{Lutz}},\ and\ \bibinfo
  {author} {\bibfnamefont{A.~J.}\ \bibnamefont{Heinrich}},\ }%
  \bibfield{journal}{%
  \bibinfo {journal} {Science}\ }%
  \textbf{\bibinfo {volume} {312}},\ \bibinfo {pages} {1021} (\bibinfo {year}
  {2006})%
  \bibAnnoteFile{NoStop}{HirjibehedinScience06}%
\bibitem{BalashovPRL06}%
  \BibitemOpen
  \bibfield{author}{%
  \bibinfo {author} {\bibfnamefont{T.}~\bibnamefont{Balashov}}, \bibinfo
  {author} {\bibfnamefont{A.~F.}\ \bibnamefont{Tak\'{a}cs}}, \bibinfo {author}
  {\bibfnamefont{W.}~\bibnamefont{Wulfhekel}},\ and\ \bibinfo {author}
  {\bibnamefont{J.Kirschner}},\ }%
  \bibfield{journal}{%
  \bibinfo {journal} {Phys. Rev. Lett.}\ }%
  \textbf{\bibinfo {volume} {97}},\ \bibinfo {pages} {187201} (\bibinfo {year}
  {2006})%
  \bibAnnoteFile{NoStop}{BalashovPRL06}%
\bibitem{RossierPRL09}%
  \BibitemOpen
  \bibfield{author}{%
  \bibinfo {author} {\bibfnamefont{J.}~\bibnamefont{Fern\'{a}ndez-Rossier}},\
  }%
  \bibfield{journal}{%
  \bibinfo {journal} {Phys. Rev. Lett.}\ }%
  \textbf{\bibinfo {volume} {102}},\ \bibinfo {pages} {256802} (\bibinfo {year}
  {2009})%
  \bibAnnoteFile{NoStop}{RossierPRL09}%
\bibitem{FranssonNanoLett09}%
  \BibitemOpen
  \bibfield{author}{%
  \bibinfo {author} {\bibfnamefont{J.}~\bibnamefont{Fransson}},\ }%
  \bibfield{journal}{%
  \bibinfo {journal} {Nano Lett.}\ }%
  \textbf{\bibinfo {volume} {9}},\ \bibinfo {pages} {2414} (\bibinfo {year}
  {2009})%
  \bibAnnoteFile{NoStop}{FranssonNanoLett09}%
\bibitem{AppelbaumPR67}%
  \BibitemOpen
  \bibfield{author}{%
  \bibinfo {author} {\bibfnamefont{J.~A.}\ \bibnamefont{Appelbaum}},\ }%
  \bibfield{journal}{%
  \bibinfo {journal} {Phys. Rev.}\ }%
  \textbf{\bibinfo {volume} {154}},\ \bibinfo {pages} {633} (\bibinfo {year}
  {1967})%
  \bibAnnoteFile{NoStop}{AppelbaumPR67}%
\bibitem{Benderbook}%
  \BibitemOpen
  \bibfield{author}{%
  \bibinfo {author} {\bibfnamefont{C.~M.}\ \bibnamefont{Bender}}\ and\ \bibinfo
  {author} {\bibfnamefont{S.~A.}\ \bibnamefont{Orszag}},\ }%
  \emph{\bibinfo {title} {Advanced Mathematical Methods for Scientists and
  Engineers: Asymptotic Methods and Perturbation Theory}}\ (\bibinfo
  {publisher} {Springer-Verlag New York},\ \bibinfo {year} {1999})%
  \bibAnnoteFile{NoStop}{Benderbook}%
\bibitem{FranssonPRB01}%
  \BibitemOpen
  \bibfield{author}{%
  \bibinfo {author} {\bibfnamefont{J.}~\bibnamefont{Fransson}}, \bibinfo
  {author} {\bibfnamefont{O.}~\bibnamefont{Eriksson}},\ and\ \bibinfo {author}
  {\bibfnamefont{I.}~\bibnamefont{Sandalov}},\ }%
  \bibfield{journal}{%
  \bibinfo {journal} {Phys. Rev. B}\ }%
  \textbf{\bibinfo {volume} {64}} (\bibinfo {year} {2001})%
  \bibAnnoteFile{NoStop}{FranssonPRB01}%
\bibitem{Note1}%
  \BibitemOpen
  \bibinfo {note} {For instance, in the extended Shubin-Hubbard model similar
  correlated hopping terms are included. See for example Ref. [\protect
  \rev@citealpnum {JPhysCVonsovsky79}] and references therein.}%
  \bibAnnoteFile{Stop}{Note1}%
\bibitem{Mahanbook}%
  \BibitemOpen
  \bibfield{author}{%
  \bibinfo {author} {\bibfnamefont{G.}~\bibnamefont{Mahan}},\ }%
  \emph{\bibinfo {title} {Many Particle Physics (Physics of Solids and
  Liquids)}},\ \bibinfo {edition} {3rd}\ ed.\ (\bibinfo {publisher} {Plenum
  Publishers},\ \bibinfo {year} {2000})%
  \bibAnnoteFile{NoStop}{Mahanbook}%
\bibitem{ThomasPRL96}%
  \BibitemOpen
  \bibfield{author}{%
  \bibinfo {author} {\bibfnamefont{K.~J.}\ \bibnamefont{Thomas}}, \bibinfo
  {author} {\bibfnamefont{J.~T.}\ \bibnamefont{Nicholls}}, \bibinfo {author}
  {\bibfnamefont{M.~Y.}\ \bibnamefont{Simmons}}, \bibinfo {author}
  {\bibfnamefont{M.}~\bibnamefont{Pepper}}, \bibinfo {author}
  {\bibfnamefont{D.~R.}\ \bibnamefont{Mace}},\ and\ \bibinfo {author}
  {\bibfnamefont{D.~A.}\ \bibnamefont{Ritchie}},\ }%
  \bibfield{journal}{%
  \bibinfo {journal} {Phys. Rev. Lett.}\ }%
  \textbf{\bibinfo {volume} {77}},\ \bibinfo {pages} {135} (\bibinfo {year}
  {1996})%
  \bibAnnoteFile{NoStop}{ThomasPRL96}%
\bibitem{MeirPRL02}%
  \BibitemOpen
  \bibfield{author}{%
  \bibinfo {author} {\bibfnamefont{Y.}~\bibnamefont{Meir}}, \bibinfo {author}
  {\bibfnamefont{K.}~\bibnamefont{Hirose}},\ and\ \bibinfo {author}
  {\bibfnamefont{N.~S.}\ \bibnamefont{Wingreen}},\ }%
  \bibfield{journal}{%
  \bibinfo {journal} {Phys. Rev. Lett.}\ }%
  \textbf{\bibinfo {volume} {89}},\ \bibinfo {pages} {196802} (\bibinfo {year}
  {2002})%
  \bibAnnoteFile{NoStop}{MeirPRL02}%
\bibitem{MeirPRL92}%
  \BibitemOpen
  \bibfield{author}{%
  \bibinfo {author} {\bibfnamefont{Y.}~\bibnamefont{Meir}}\ and\ \bibinfo
  {author} {\bibfnamefont{N.~S.}\ \bibnamefont{Wingreen}},\ }%
  \bibfield{journal}{%
  \bibinfo {journal} {Phys. Rev. Lett.}\ }%
  \textbf{\bibinfo {volume} {68}},\ \bibinfo {pages} {2512} (\bibinfo {year}
  {1992})%
  \bibAnnoteFile{NoStop}{MeirPRL92}%
\bibitem{Haugbook}%
  \BibitemOpen
  \bibfield{author}{%
  \bibinfo {author} {\bibfnamefont{H.}~\bibnamefont{Haug}}\ and\ \bibinfo
  {author} {\bibfnamefont{A.}~\bibnamefont{Jauho}},\ }%
  \emph{\bibinfo {title} {Quantum Kinetics in Transport and Optics of
  Semiconductors}}\ (\bibinfo {publisher} {Springer},\ \bibinfo {year} {1998})%
  \bibAnnoteFile{NoStop}{Haugbook}%
\bibitem{JPhysCVonsovsky79}%
  \BibitemOpen
  \bibfield{author}{%
  \bibinfo {author} {\bibfnamefont{S.~V.}\ \bibnamefont{Vonsovsky}}\ and\
  \bibinfo {author} {\bibfnamefont{M.~I.}\ \bibnamefont{Katsnelson}},\ }%
  \bibfield{journal}{%
  \bibinfo {journal} {J. Phys. C: Solid State Phys.}\ }%
  \textbf{\bibinfo {volume} {12}},\ \bibinfo {pages} {2043} (\bibinfo {year}
  {1979})%
  \bibAnnoteFile{NoStop}{JPhysCVonsovsky79}%
\end{thebibliography}%

\end{document}